\documentclass[10pt]{article}
\usepackage[T1]{fontenc}
\usepackage[utf8]{inputenc}
\usepackage{graphicx}
\usepackage{listings}
\usepackage{bm}
\usepackage{latexsym}
\usepackage{amsmath}
\usepackage{booktabs}
\usepackage{array}
\usepackage[UKenglish]{babel}
\usepackage[a4paper,top=1.5cm,bottom=1.5cm,left=1.5cm,right=1.5cm]{geometry}
\usepackage{lineno}
\usepackage{tabularx}
\usepackage{longtable}
\RequirePackage[colorlinks=true
,urlcolor=blue
,anchorcolor=blue
,citecolor=blue
,filecolor=blue
,linkcolor=blue
,menucolor=blue
,pagecolor=blue
,linktocpage=true
,pdfproducer=medialab
]{hyperref}

\usepackage{makeidx}
\newcommand{\be}{\begin{equation}}  
\newcommand{\ee}{\end{equation}}  
\newcommand{\bea}{\begin{eqnarray}}  
\newcommand{\eea}{\end{eqnarray}}  
\makeindex

\makeatletter
\g@addto@macro\bfseries{\boldmath}
\makeatother

\newcommand{\MYhref}[3][blue]{\href{#2}{\color{#1}{#3}}}

\begin{document}


\begin{center}
{\LARGE \bf The ATLASpdf21 fit: a novel determination of proton Parton Distribution Functions using ATLAS data}

\par\vspace*{2.5mm}\par

{

\bigskip

\large \bf Francesco Giuli\footnote{E-Mail: \MYhref{francesco.giuli@cern.ch}{francesco.giuli@cern.ch}} (on behalf of the ATLAS Collaboration\footnote{Copyright 2022 CERN for the benefit of the ATLAS Collaboration. Reproduction of this article or parts of it is allowed as specified in the CC-BY-4.0 license.})}

\vspace*{2.5mm}

{CERN, EP Department, CH-1211 Geneva 23, Switzerland}

\vspace*{2.5mm}

{\it Presented at DIS2022: XXIX International Workshop on Deep-Inelastic Scattering and Related Subjects, Santiago de Compostela, Spain, May 2-6 2022}

\vspace*{2.5mm}

\end{center}

\begin{abstract}

We present fits to determine Parton Distribution Functions using a diverse set of measurements from the ATLAS experiment at the LHC, including inclusive $W$ and $Z$ boson production, $t\bar{t}$ production, $W$+jets and $Z$+jets production, inclusive jet production and direct photon production. These ATLAS measurements are used in combination with deep-inelastic scattering data from the electron-proton collider HERA. Particular attention is paid to the correlation of systematic uncertainties within and between the various ATLAS data sets and to the impact of model, theoretical and parameterisation uncertainties.
\end{abstract}
 
\section{Introduction and input data sets}
Parton Distribution Functions (PDFs) represent one of the main sources of theoretical systematic uncertainties in hadronic collisions. A precise knowledge of PDFs is a necessary ingredient for accurate predictions of Standard Model (SM). Furthermore, they influence the potential of experimental searches for discovering or setting exclusion bounds in many Beyond the Standard Model (BSM) scenarios.\\
In this proceeding, a review of the most recent determination of PDFs by the ATLAS Collaboration~\cite{ATLAS:2021vod} is presented. This fit represents the first comprehensive and comparative quantum chromodynamics (QCD) analysis of a number of ATLAS data sets with potential sensitivity to PDFs. The data sets in use are described in the following: the high precision measurements of the inclusive differential $W^{\pm}$ and $Z/\gamma^{*}$ boson cross sections at 7 TeV~\cite{ATLAS:2016nqi}, differential $t\bar{t}$ distributions in the lepton + jets and dilepton channels at 8 TeV~\cite{ATLAS:2018owm}, data on the production of $W$ and $Z$ bosons in association with jets ($V$ + jets)~\cite{ATLAS:2017irc,ATLAS:2019bsa}, the data on $W$~\cite{ATLAS:2019fgb} production and $Z/\gamma^{*}$~\cite{ATLAS:2017rue} production with 20.2 fb$^{-1}$ at 8 TeV, the direct photon production differential cross sections with 20.2 fb$^{-1}$ at 8 TeV and 3.2 fb$^{-1}$ at 13 TeV are added in the form of their ratios~\cite{ATLAS:2019drj}, the $t\bar{t}$ differential cross sections in the lepton + jets channel from 3.2 fb$^{-1}$ at 13 TeV~\cite{ATLAS:2019hxz} and inclusive jet production cross sections with 4.5 fb$^{-1}$ at 7 TeV~\cite{ATLAS:2014riz}, 20.2 fb$^{-1}$ at 8 TeV~\cite{ATLAS:2017kux} and 3.2 fb$^{-1}$ at 13 TeV~\cite{ATLAS:2017ble}. All these analyses use $pp$ collisions data collected with the ATLAS detector~\cite{ATLAS:2008xda} at the Large Hadron Collider (LHC).\\
The ATLAS data sets are added on top of the combined $e^{\pm}p$ cross-section measurements of H1 and ZEUS~\cite{H1:2015ubc}. These HERA data are the backbone of modern QCD analyses because they cover a kinematic range of $Q^{2}$ from 4.5 $\cdot$ 10$^{-2}$~GeV$^{2}$ to 5 $\cdot$ 10$^{4}$~GeV$^{2}$ and for Bjorken-$x$ from 6 $\cdot$ 10$^{-7}$ to 0.65 (Neutral Current - NC) and Q2 $\sim$ 300~GeV$^{2}$ to beyond 10$^{4}$~GeV$^{2}$ and of $x$ from $\sim$ 0.65 down to $\sim$ 10$^{-2}$ (Charged Current - CC).

\section{Theoretical framework and fit methodology}
The \texttt{xFitter} framework~\cite{Alekhin:2014irh,H1:2009pze,H1:2009bcq} has been interfaced to theoretical calculations directly or uses fast interpolation grids to make theoretical predictions for the considered processes. The predictions for all the processes are calculated to next-to-next-to-leading order (NNLO) in QCD and next-to-leading order (NLO) in electroweak (EW) theory. The minimisation is performed using \texttt{MINUIT}~\cite{James:1975dr}. PDFs are parametrised as a function of $x$ at a starting scale $Q_{0}^{2}$ = 1.9~GeV$^{2}$. The values of the heavy-quark masses are $m_{c}$ = 1.41~GeV and $m_{b}$ = 4.2~GeV respectively, as suggested in a recent re-analysis of the HERA heavy-quark data~\cite{H1:2021xxi}. A cut on the minimum $Q^{2}$ of the data to enter the fit is imposed at $Q^{2}_{\mathrm{min}}$ = 10.0~GeV$^{2}$. All the fits are performed with a fixed value of the strong coupling constant set to $\alpha_{S}(m_{Z})$ = 0.118.\\
The quark distributions are parametrised at $Q_{0}$ by the generic functional form $xq_{i}(x) = A_{i}x^{B_{i}}(1-x)^{C_{i}}P_{i}(x)$, while the gluon PDF is represented by the more flexible form
$xg(x) = A_{g}x^{B_{g}}(1-x)^{C_{g}}P_{g}(x)-A_{g}^{'}x^{B^{'}_{g}}(1-x)^{C^{'}_{g}}$, where $C^{'}_{g}$ = 25 to avoid a negative gluon at high $x$, and $P_{i,g}(x)=(1+D_{i,g}x+E_{i,g}x^{2}+F_{i,g}x^{3})$ for both quarks and gluon distributions. The parametrised distributions at the starting scale are chosen to be $xu_{V}$, $xd_{V}$, $x\bar{u}$, $x\bar{d}$, $x\bar{s}$ and $xg$. The $D$, $E$ and $F$ terms in the polynomial expansion $P_{i,g}(x)$ are added only if the $\chi^{2}$ of the fit decreases significantly, till the so-called point of `saturation' of the $\chi^{2}$ is reached. This procedure leads to a central fit with 21 free parameters. We refer to the resulting set of parton distribution functions as `ATLASpdf21' in the following.\\
The level of agreement between the predictions from a PDF parametrisation and the data is quantified by the $\chi^{2}$ per degree of freedom ($\chi^{2}$/NDF), following the definition given in Ref.~\cite{H1:2015ubc}.
The experimental uncertainties are first set using the usual tolerance $T$ = 1, where $T^{2}=\Delta\chi^{2}=1$. The use of an enhanced tolerance is considered in Section~\ref{sec:tolerance}.

\section{Inclusion of correlations between data sets}
\begin{figure}[t!]
\begin{center}
\includegraphics[width=0.443\textwidth]{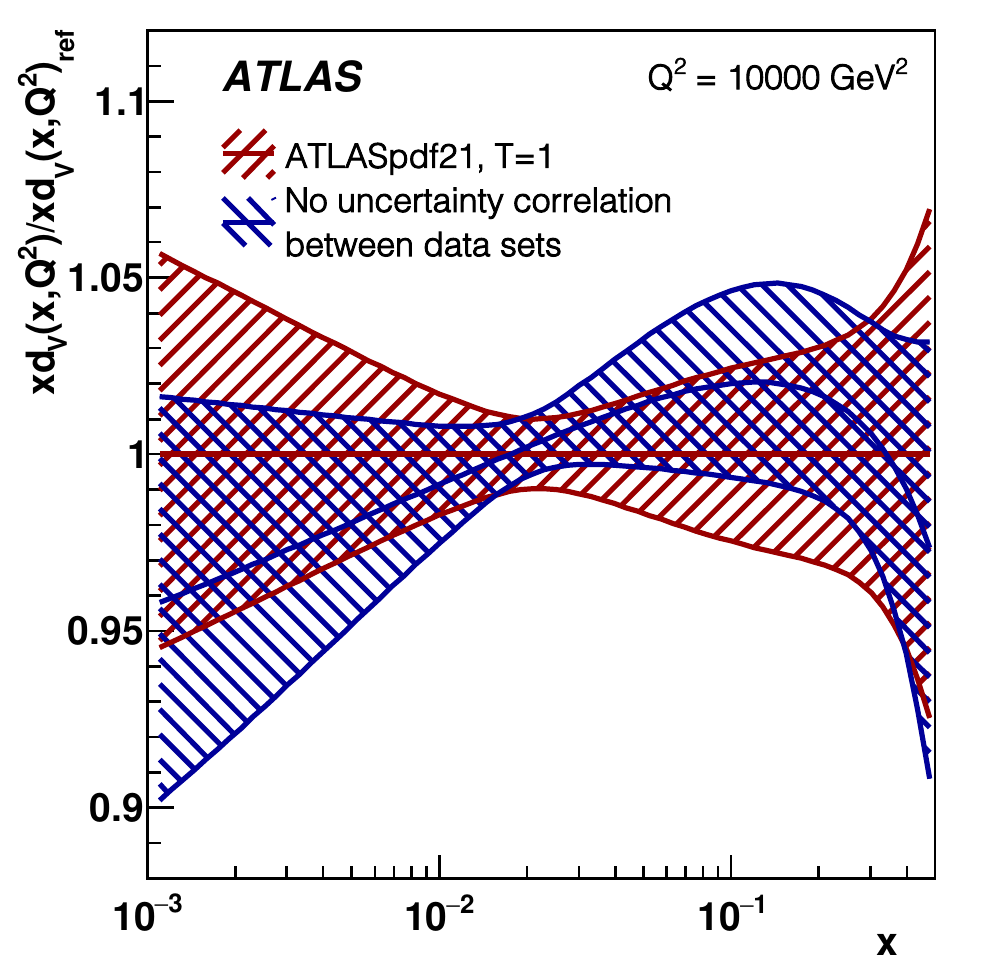}
\includegraphics[width=0.443\textwidth]{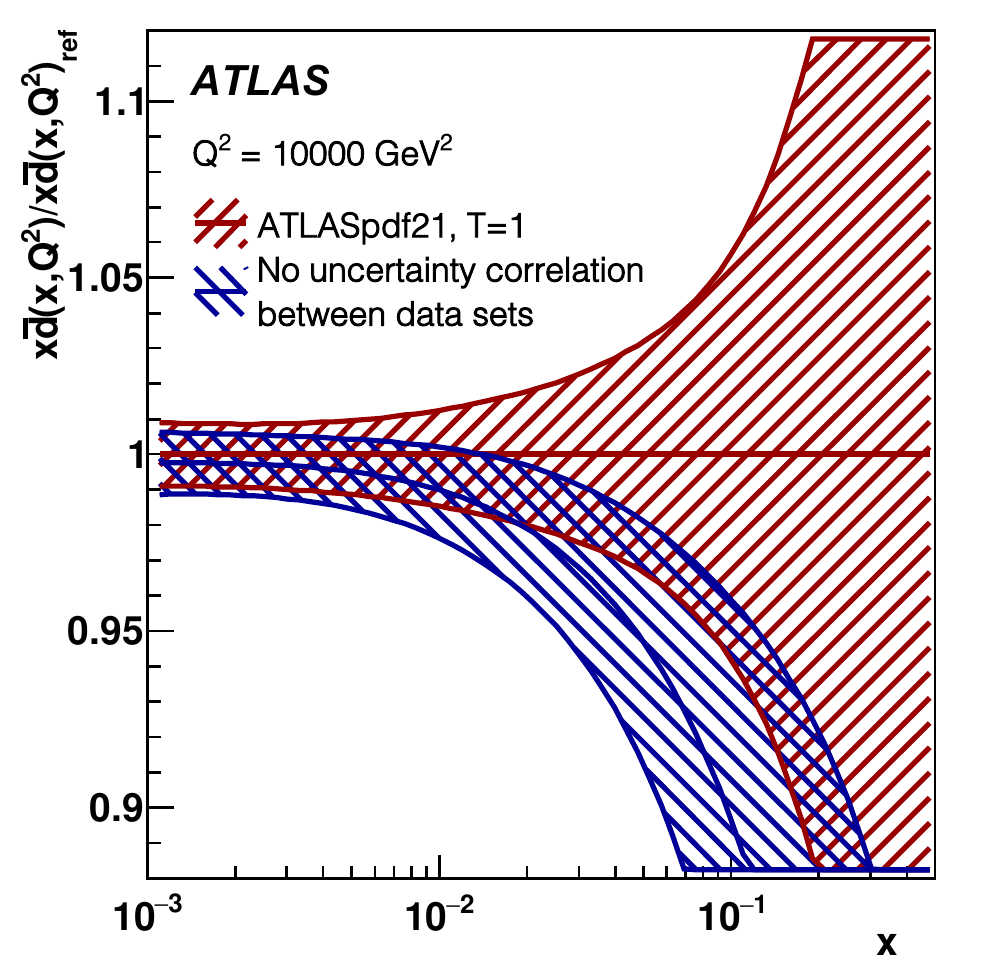}
\end{center}
\caption{ATLASpdf21 PDFs at $Q^{2}$ = 10$^{4}$~GeV$^{2}$ comparing those extracted from a fit in which correlations of systematic uncertainties between data sets are applied, with those extracted from a fit in which only the luminosity uncertainties for each centre-of-mass energy are correlated between data sets. Experimental uncertainties are shown, evaluated with tolerance $T$ = 1. Left: $xd_{V}$. Right: $x\bar{d}$. These plots are taken from Ref.~\cite{ATLAS:2021vod}.} 
\label{fig:corr}
\end{figure}

In Ref.~\cite{ATLAS:2021vod}, the correlation of systematic uncertainties within each data set and between data sets has been carefully investigated. The differences between the extracted PDFs with and without inter-data-set correlations are mostly in the $d$-type sector and can reach $\sim$ 20$\%$ for $x\bar{d}$ at large $x$, as illustrated in Figure~\ref{fig:corr} at  scale relevant for precision LHC physics, namely $Q^2$ = 10$^{4}$~GeV$^{2}$. Although these differences are not large compared to current experimental precision, they can nevertheless be important if the desired accuracy of the PDFs is $\mathcal{O}$(1$\%$), as stated in a recent NNPDF stduy~\cite{Ball:2021leu}.
\begin{figure}[t!]
\begin{center}
\includegraphics[width=0.443\textwidth]{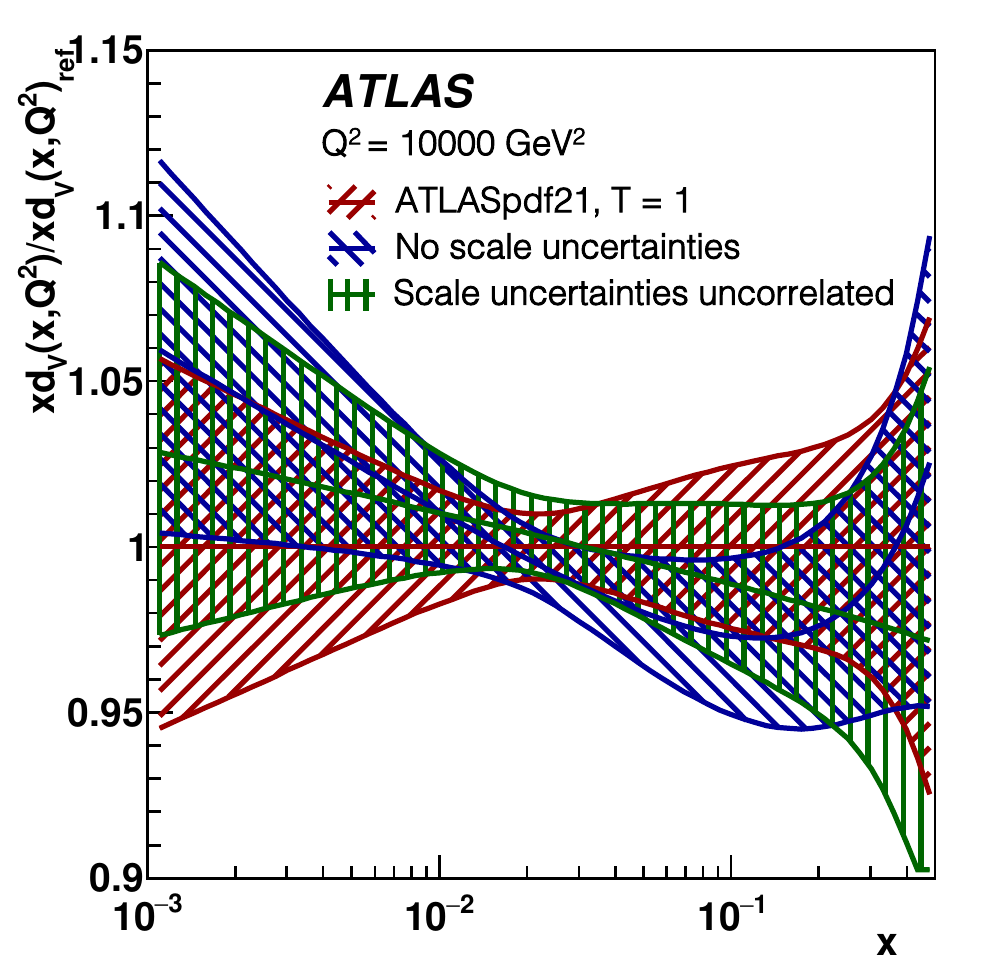}
\includegraphics[width=0.443\textwidth]{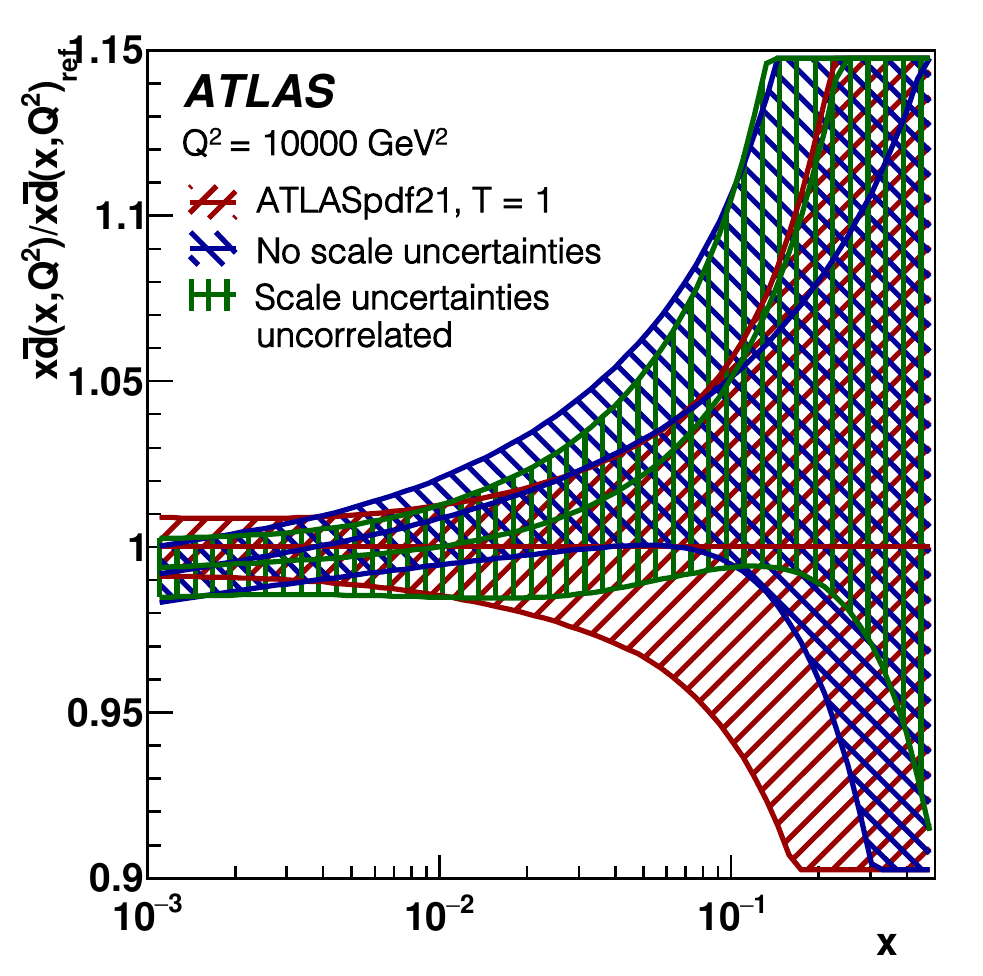}
\end{center}
\caption{ATLASpdf21, showing the ratios of a fit not including theoretical scale uncertainties in the inclusive $W$, $Z$ data to the central fit which does include these uncertainties, at the scale $Q^{2}$ = 10$^{4}$ GeV$^{2}$. Experimental uncertainties are shown, evaluated with tolerance $T$ = 1. Left: $xd_{V}$. Right: $x\bar{d}$. These plots are taken from Ref.~\cite{ATLAS:2021vod}.} 
\label{fig:scale}
\end{figure}

\section{Inclusion of scale uncertainties}
\begin{figure}[t!]
\begin{center}
\includegraphics[width=0.443\textwidth]{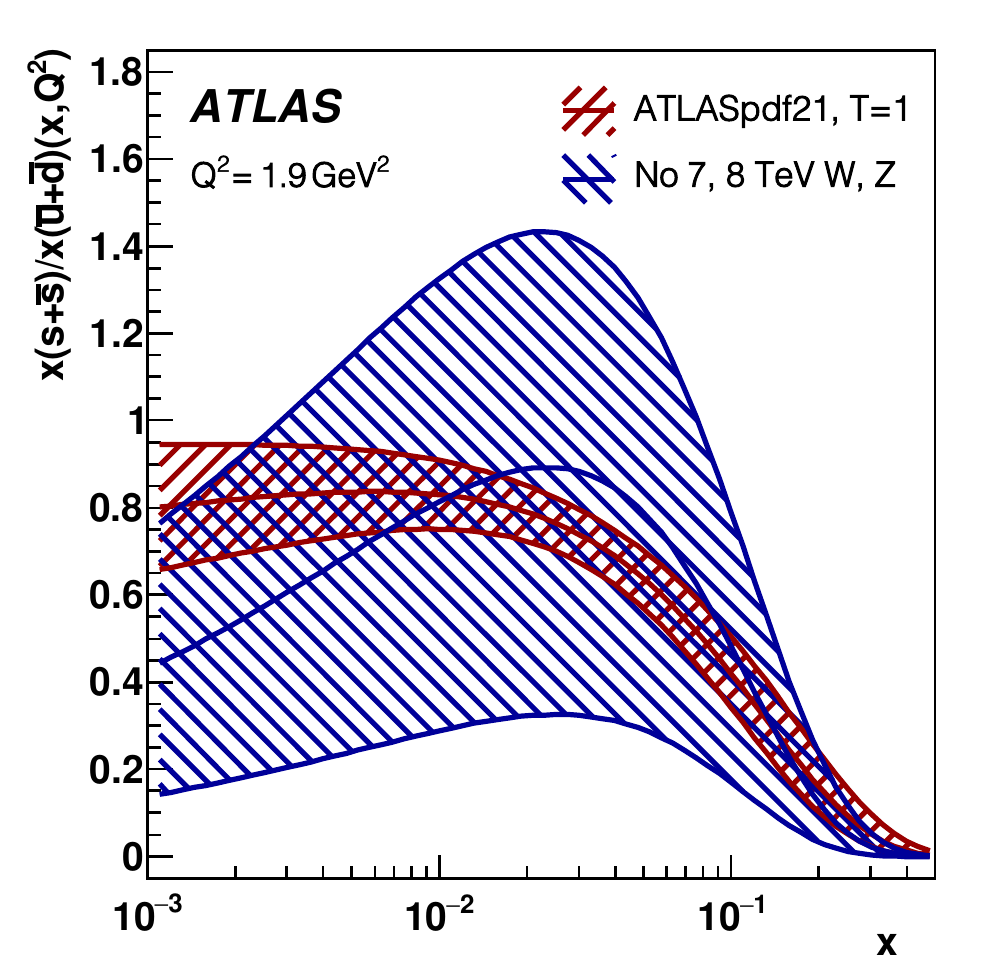}
\includegraphics[width=0.443\textwidth]{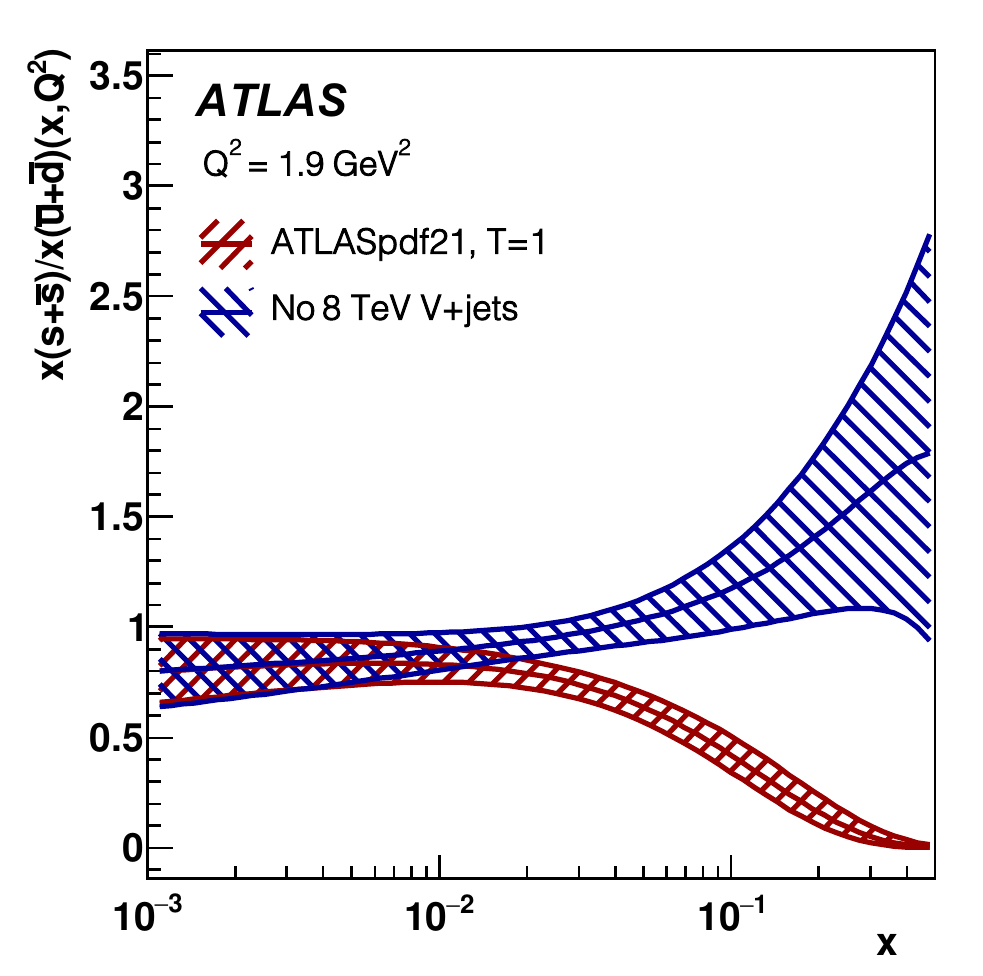}
\includegraphics[width=0.443\textwidth]{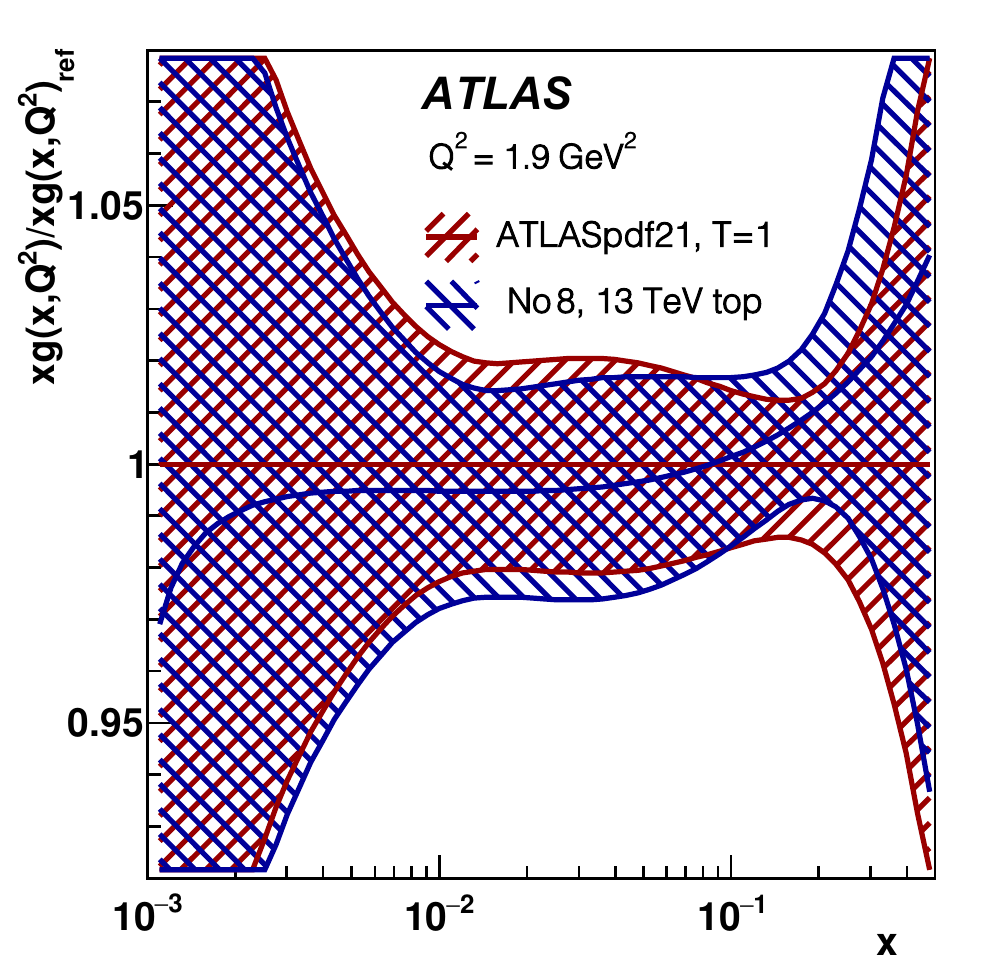}
\includegraphics[width=0.443\textwidth]{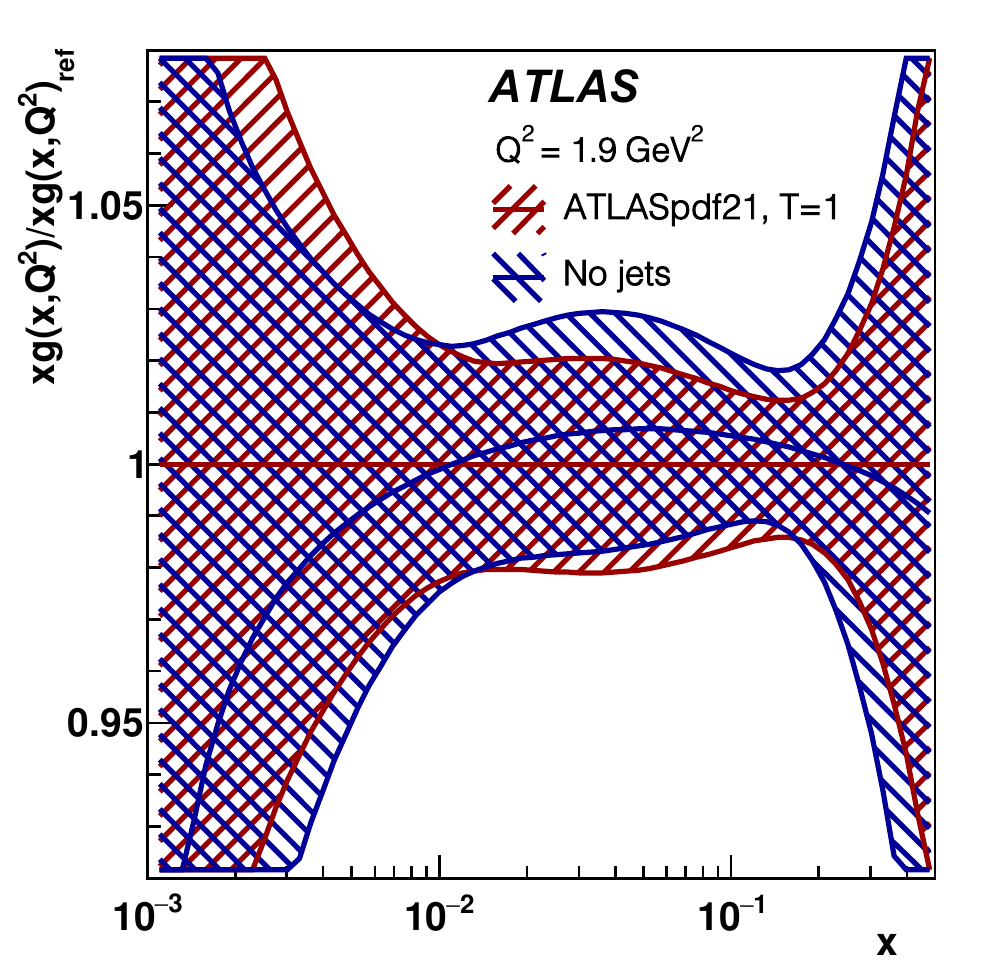}
\end{center}
\caption{ATLASpdf21 PDFs compared with those from a fit not including various data sets. Experimental uncertainties are shown, evaluated with tolerance $T$ = 1. Top left: without $W$, $Z$ data at 7 and 8~TeV ($R_{s}=x(s+\bar{s})/x(\bar{u}+\bar{d})$). Top right: without $V$ + jets data at 8~TeV ($R_{s}=x(s+\bar{s})/x(\bar{u}+\bar{d})$). Bottom left: without $t\bar{t}$ data at 8 and 13~TeV ($xg$). Bottom left: without inclusive jet data at 8~TeV ($xg$). These plots are taken from Ref.~\cite{ATLAS:2021vod}.} 
\label{fig:impact}
\end{figure}
Only for the data sets where the precision of the measurements is comparable to the size of the NNLO scale uncertainties, we considered the inclusion of these uncertainties as additional theoretical uncertainties. These data sets are the ATLAS $W$ and $Z/\gamma^{*}$ inclusive ones at both 7 and 8~TeV. Due to the similarity of the $W$ and $Z$ processes, both the factorisation ($\mu_{F}$) and renormalisation ($\mu_{R}$) scales are treated as correlated between these two data set. Furthermore, alternative scenarios where scale uncertainties are included but not correlated between 7 and 8~TeV data sets or where they are not applied at all are considered as well. Figure~\ref{fig:scale} shows the results of the two above-mentioned cases, compared with the central fit, at a scale $Q^{2}$ = 10$^{4}$ GeV$^{2}$. The differences between the shapes of the PDFs are not large, but they can be important for precise PDF determination. The size of the PDF uncertainties is very similar, whether or not the scale uncertainties are included. For the other processes the experimental uncertainties are larger than the scale uncertainties and in this case these uncertainties are treated by repeating the fit with varied scales.

\section{Impact of the various data sets on PDFs}
Figure~\ref{fig:impact} shows the $R_{s}=x(s+\bar{s})/x(\bar{u}+\bar{d})$ for the ATLASpdf21 fit compared with a fit without the inclusion of $W$, $Z$ data at 7 and 8~TeV. This ratio cannot be determined reliably without $W$, $Z$ inclusive data. Adding the $V$ + jets data at 8 TeV changes the $x\bar{d}$ and $x\bar{s}$ PDF shapes at high $x$ and resolves a double minimum in parameter space such that a harder $x\bar{d}$ and a softer $x\bar{s}$ are now preferred at high $x$ (a similar result has been already discussed in the ATLASepWZVjets20 PDF fit~\cite{ATLAS:2021qnl}). $t\bar{t}$ data soften the $xg$ PDF at high $x$ and reduce the uncertainties for $x$ > 0.1 by $\sim$ 20$\%$, as shown in Figure~\ref{fig:impact}. The principal impact of the inclusive jet data is on the gluon PDF. They prefer a mildly harder gluon at high $x$, as shown in Figure~\ref{fig:impact} and the addition of this specific data set decreases the PDF uncertainties for $x$ > 0.2 considerably.
\begin{figure}[t!]
\begin{center}
\includegraphics[width=0.443\textwidth]{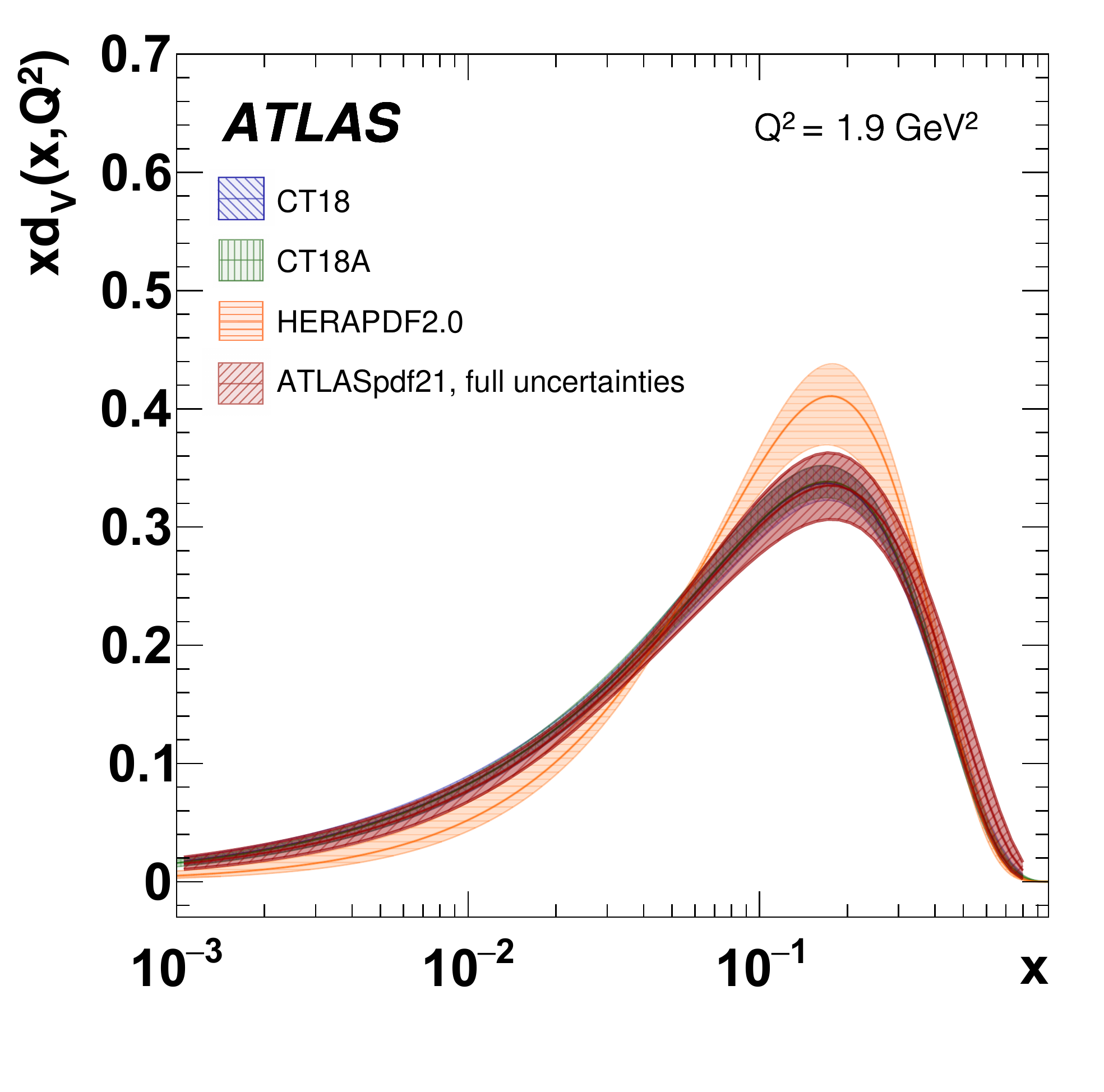}
\includegraphics[width=0.443\textwidth]{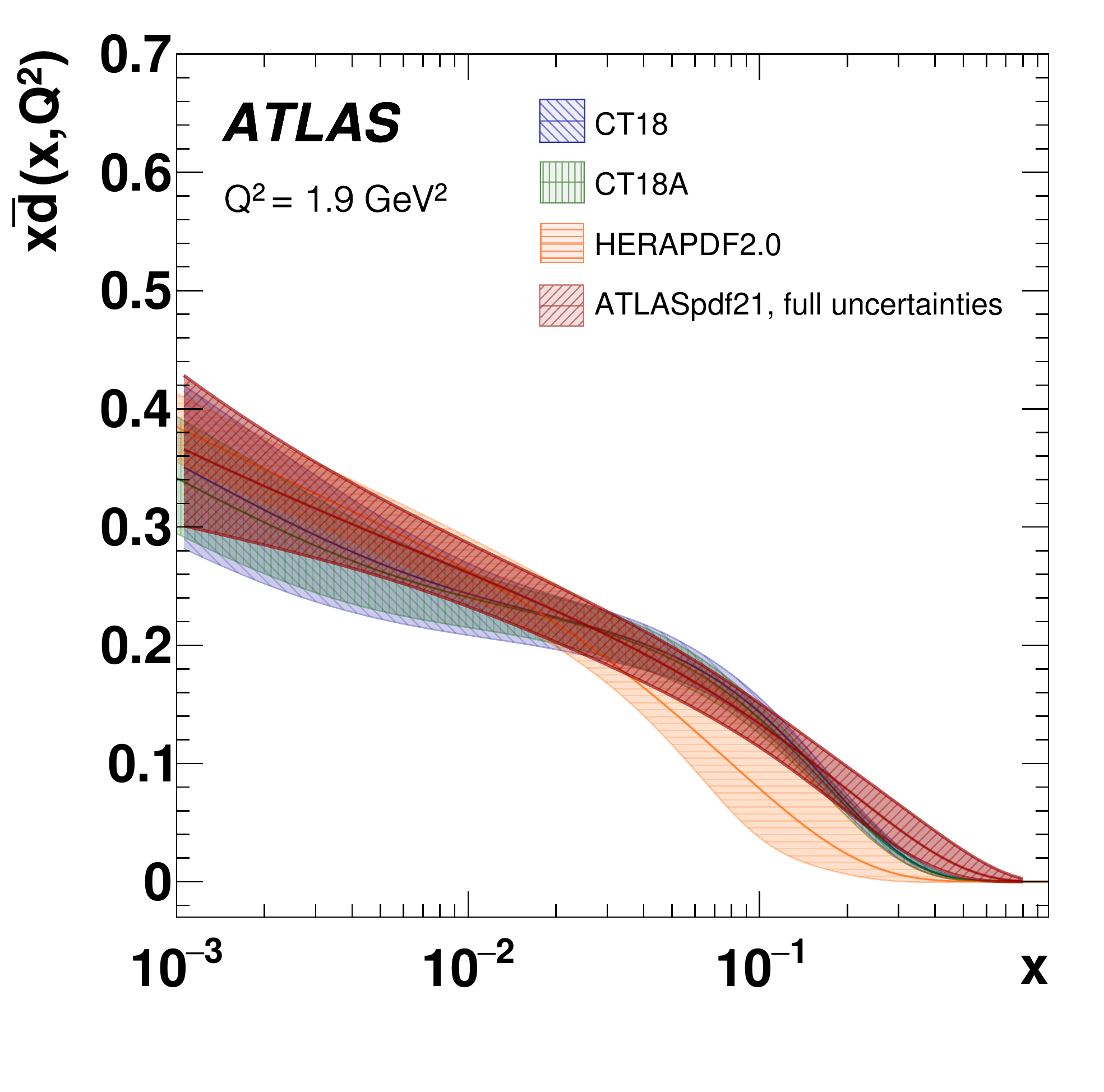}
\includegraphics[width=0.443\textwidth]{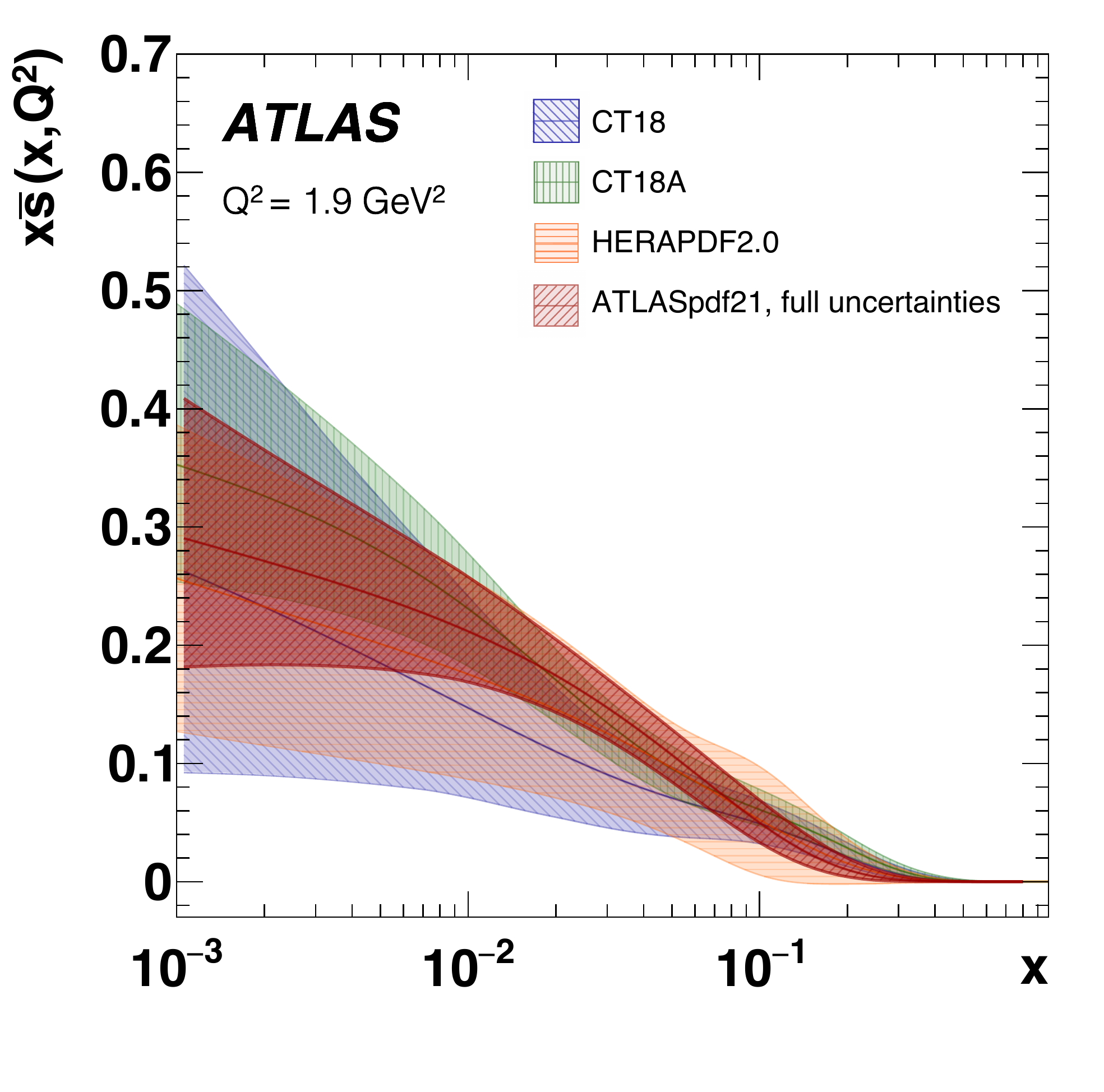}
\includegraphics[width=0.443\textwidth]{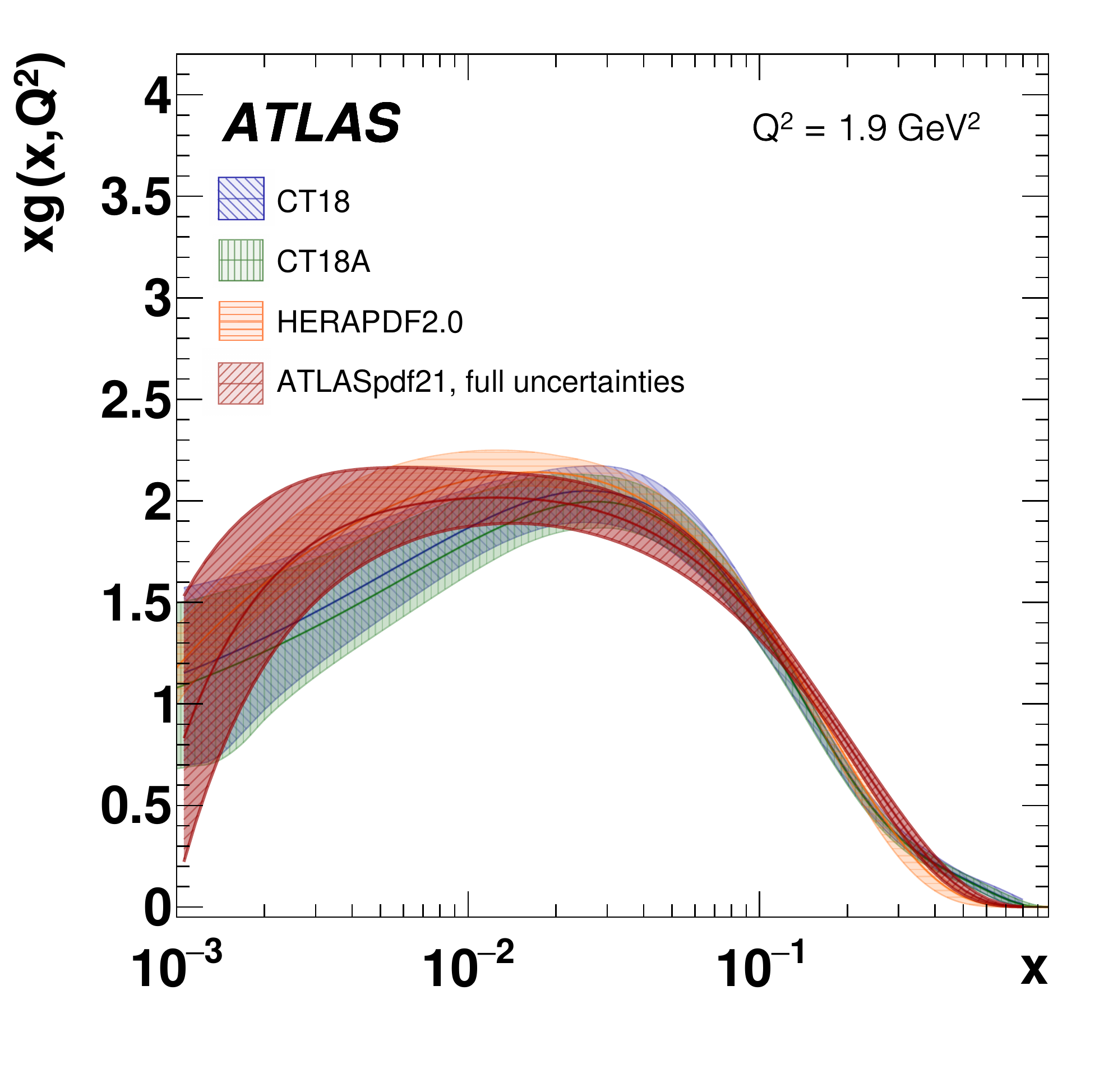}
\end{center}
\caption{ATLASpdf21 distributions with full uncertainties evaluated with $T$ = 3 compared with CT18, CT18A and HERAPDF2.0. Top left: $xd_{V}$. Tor right: $x\bar{d}$. Bottom left: $x\bar{s}$. Bottom right: $xg$. These plots are taken from Ref.~\cite{ATLAS:2021vod}.} 
\label{fig:PDFs}
\end{figure}

\section{Enhanced tolerance and comparison with modern PDF fits}
\label{sec:tolerance}
To take into account possible inconsistencies between the independent fitted data sets and unknown experimental and theoretical uncertainties, following the dynamic tolerance procedure introduced in the MSTW paper~\cite{Martin:2009iq}, a tolerance of $T$ = 3 is found to be a reasonable choice and is applied for the final estimation of PDF uncertainty.\\
The ATLASpdf21 PDFs evaluated with full uncertainties are compared to HERAPDF2.0~\cite{H1:2015ubc}, CT18~\cite{Hou:2019efy} and CT18A~\cite{Hou:2019efy} in Figure~\ref{fig:PDFs}. Looking at the $xd_{V}$ and $x\bar{d}$ distributions, it is clear that the ATLASpdf21 PDFs get closer to global fits, as soon as more LHC data sets are included. It is very nice to notice that the $x\bar{s}$ distribution is in good agreement with CT18A (the only fit which includes the ATLAS $W$ and $Z/\gamma^{*}$ data at 7 TeV). Finally, the gluon distribution agrees best with HERAPDF2.0, but is in a reasonable agreement  with the other global PDFs. The ATLASpdf21 fit exhibits a $\chi^{2}/\mathrm{NDF}$ of 2010/1620, while, for the considered global fits, the $\chi^{2}$ values for the data sets included in the ATLASpdf21 fit are HERAPDF2.0: 2262, CT18: 2135 and CT18A: 2133. Interestingly, the $\chi^{2}$ value of the ATLASpdf21 fit is better for the considered data sets, although the global PDFs from CT have more flexible parametrisation.

\section{Conclusion}
In this proceeding, we presented the ATLASpdf21 fit, the first one by an experimental collaboration at the LHC which uses an enhanced tolerance for the computation of experimental PDF uncertainties. Furthermore, the role of scale uncertainties has been considered and these additional theoretical uncertainties have been implemented where they are most impactful. Moreover, correlation of systematic uncertainties within and between ATLAS data sets are considered. The effects of these correlations, as well as the inclusion of scale uncertainties, are relatively small a scale relevant for precision physics at the LHC, but large enough to be considered in future measurements of SM parameters and searches for BSM physics.


\begin{thebibliography}{ieetr}

\bibitem{ATLAS:2021vod}
{ATLAS Collaboration},
Eur. Phys. J. \textbf{C82} (2022) no. 5, 438
doi:10.1140/epjc/s10052-022-10217-z
[arXiv:2112.11266 [hep-ex]].

\bibitem{ATLAS:2016nqi}
{ATLAS Collaboration},
Eur. Phys. J. \textbf{C77} (2017) no. 6, 367
doi:10.1140/epjc/s10052-017-4911-9
[arXiv:1612.03016 [hep-ex]].

\bibitem{ATLAS:2018owm}
{ATLAS Collaboration},
ATL-PHYS-PUB-2018-017,
url: \MYhref{https://cds.cern.ch/record/2633819}{https://cds.cern.ch/record/2633819}.

\bibitem{ATLAS:2017irc}
{ATLAS Collaboration},
JHEP \textbf{05} (2018), 077
[erratum: JHEP \textbf{10} (2020), 048]
doi:10.1007/JHEP05(2018)077
[arXiv:1711.03296 [hep-ex]].

\bibitem{ATLAS:2019bsa}
{ATLAS Collaboration},
Eur. Phys. J. \textbf{C79} (2019) no. 10, 847
doi:10.1140/epjc/s10052-019-7321-3
[arXiv:1907.06728 [hep-ex]].

\bibitem{ATLAS:2019fgb}
{ATLAS Collaboration},
Eur. Phys. J. \textbf{C79} (2019) no. 9, 760
doi:10.1140/epjc/s10052-019-7199-0
[arXiv:1904.05631 [hep-ex]].

\bibitem{ATLAS:2017rue}
{ATLAS Collaboration},
JHEP \textbf{12} (2017), 059
doi:10.1007/JHEP12(2017)059
[arXiv:1710.05167 [hep-ex]].

\bibitem{ATLAS:2019drj}
{ATLAS Collaboration},
JHEP \textbf{04} (2019), 093
doi:10.1007/JHEP04(2019)093
[arXiv:1901.10075 [hep-ex]].

\bibitem{ATLAS:2019hxz}
{ATLAS Collaboration},
Eur. Phys. J. \textbf{C79} (2019) no. 12, 1028
[erratum: Eur. Phys. J. \textbf{C80} (2020) no. 11, 1092]
doi:10.1140/epjc/s10052-019-7525-6
[arXiv:1908.07305 [hep-ex]].

\bibitem{ATLAS:2014riz}
{ATLAS Collaboration},
JHEP \textbf{02} (2015), 153
[erratum: JHEP \textbf{09} (2015), 141]
doi:10.1007/JHEP02(2015)153
[arXiv:1410.8857 [hep-ex]].

\bibitem{ATLAS:2017kux}
{ATLAS Collaboration},
JHEP \textbf{09} (2017), 020
doi:10.1007/JHEP09(2017)020
[arXiv:1706.03192 [hep-ex]].

\bibitem{ATLAS:2017ble}
{ATLAS Collaboration},
JHEP \textbf{05}, 195 (2018)
doi:10.1007/JHEP05(2018)195
[arXiv:1711.02692 [hep-ex]].

\bibitem{ATLAS:2008xda}
{ATLAS Collaboration},
JINST \textbf{3} (2008), S08003
doi:10.1088/1748-0221/3/08/S08003

\bibitem{H1:2015ubc}
{H1 and ZEUS Collaborations},
Eur. Phys. J. \textbf{C75} (2015) no. 12, 580
doi:10.1140/epjc/s10052-015-3710-4
[arXiv:1506.06042 [hep-ex]].

\bibitem{Alekhin:2014irh}
S.~Alekhin \textit{et al.},
Eur. Phys. J. \textbf{C75} (2015) no. 7, 304
doi:10.1140/epjc/s10052-015-3480-z
[arXiv:1410.4412 [hep-ph]].

\bibitem{H1:2009pze}
{H1 and ZEUS Collaborations},
JHEP \textbf{01} (2010), 109
doi:10.1007/JHEP01(2010)109
[arXiv:0911.0884 [hep-ex]].

\bibitem{H1:2009bcq}
{H1 Collaboration},
Eur. Phys. J. \textbf{C64} (2009), 561-587
doi:10.1140/epjc/s10052-009-1169-x
[arXiv:0904.3513 [hep-ex]].

\bibitem{James:1975dr}
F.~James and M.~Roos,
Comput. Phys. Commun. \textbf{10} (1975), 343-367
doi:10.1016/0010-4655(75)90039-9

\bibitem{H1:2021xxi}
{H1 and ZEUS Collaborations},
Eur. Phys. J. \textbf{C82} (2022) no. 3, 243
doi:10.1140/epjc/s10052-022-10083-9
[arXiv:2112.01120 [hep-ex]].

\bibitem{Ball:2021leu}
{NNPDF Collaboration}
Eur. Phys. J. \textbf{C82} (2022) no. 5, 428
doi:10.1140/epjc/s10052-022-10328-7
[arXiv:2109.02653 [hep-ph]].

\bibitem{ATLAS:2021qnl}
{ATLAS Collaboration}
JHEP \textbf{07} (2021), 223
doi:10.1007/JHEP07(2021)223
[arXiv:2101.05095 [hep-ex]].

\bibitem{Martin:2009iq}
A.~D.~Martin \textit{et al.},
Eur. Phys. J. C \textbf{63} (2009), 189-285
doi:10.1140/epjc/s10052-009-1072-5
[arXiv:0901.0002 [hep-ph]].

\bibitem{Hou:2019efy}
T.~J.~Hou \textit{et al.},
Phys. Rev. D \textbf{103}, no.1, 014013 (2021)
doi:10.1103/PhysRevD.103.014013
[arXiv:1912.10053 [hep-ph]].

\end{thebibliography}
\end{document}